\pdfoutput=1
\documentclass[12pt,a4paper]{article}

\usepackage{ifthen} 
\newboolean{pdflatex}
\setboolean{pdflatex}{true} 

\newboolean{articletitles}
\setboolean{articletitles}{true} 

\newboolean{uprightparticles}
\setboolean{uprightparticles}{false} 

\def\paperauthors{LHCb collaboration} 
\def\paperasciititle{Search for the decays B(c)+ -> mu+ nu gamma} 
\def\papertitle{Search for the decays $B^+_{(c)} \to \mu^+ \nu_\mu \gamma$} 
\def\paperkeywords{{High Energy Physics}, {LHCb}} 
\def\papercopyright{\the\year\ CERN for the benefit of the LHCb collaboration} 
\def\paperlicence{CC BY 4.0 licence}
\def\paperlicenceurl{https://creativecommons.org/licenses/by/4.0/}

\newif\ifEnableSectionTOCLinks
\EnableSectionTOCLinksfalse 

\usepackage[top=1in, bottom=1.25in, left=1in, right=1in]{geometry}

\usepackage{microtype}
\usepackage{lineno}  
\usepackage{xspace} 
\usepackage{caption} 

\usepackage{graphicx}  
\usepackage{color}
\usepackage{colortbl}
\graphicspath{{./figs/}} 

\usepackage{amsmath} 
\usepackage{amssymb}
\usepackage{amsfonts}
\usepackage{upgreek} 

\newcommand*\patchAmsMathEnvironmentForLineno[1]{%
\expandafter\let\csname old#1\expandafter\endcsname\csname #1\endcsname
\expandafter\let\csname oldend#1\expandafter\endcsname\csname
end#1\endcsname
 \renewenvironment{#1}%
   {\linenomath\csname old#1\endcsname}%
   {\csname oldend#1\endcsname\endlinenomath}%
}
\newcommand*\patchBothAmsMathEnvironmentsForLineno[1]{%
  \patchAmsMathEnvironmentForLineno{#1}%
  \patchAmsMathEnvironmentForLineno{#1*}%
}
\AtBeginDocument{%
\patchBothAmsMathEnvironmentsForLineno{equation}%
\patchBothAmsMathEnvironmentsForLineno{align}%
\patchBothAmsMathEnvironmentsForLineno{flalign}%
\patchBothAmsMathEnvironmentsForLineno{alignat}%
\patchBothAmsMathEnvironmentsForLineno{gather}%
\patchBothAmsMathEnvironmentsForLineno{multline}%
\patchBothAmsMathEnvironmentsForLineno{eqnarray}%
}

\usepackage[pdftex,
            pdfauthor={\paperauthors},
            pdftitle={\paperasciititle},
            pdfkeywords={\paperkeywords}]{hyperref}
\usepackage{hyperxmp}
\hypersetup{
    pdfcopyright={Copyright (C) \papercopyright},
    pdflicenseurl={\paperlicenceurl}
}

\usepackage[colorinlistoftodos,textsize=scriptsize]{todonotes}

\usepackage[bottom,flushmargin,hang,multiple]{footmisc}

\usepackage[all]{hypcap} 

\usepackage{xspace} 
\usepackage{upgreek}

\def\lhcb   {\mbox{LHCb}\xspace}

\def\babar  {\mbox{BaBar}\xspace}
\def\belle  {\mbox{Belle}\xspace}

\def\MagUp {\mbox{\em Mag\kern -0.05em Up}\xspace}

\ifthenelse{\boolean{uprightparticles}}%
{

 \def\Peta        {\ensuremath{\upeta}\xspace}

 \def\Pmu         {\ensuremath{\upmu}\xspace}

 \def\Ppi         {\ensuremath{\uppi}\xspace}

 \def\Ppsi        {\ensuremath{\uppsi}\xspace}

 \def\PDelta      {\ensuremath{\Delta}\xspace}                 
 \def\PXi         {\ensuremath{\Xi}\xspace}                 
 \def\PLambda     {\ensuremath{\Lambda}\xspace}                 
 \def\PSigma      {\ensuremath{\Sigma}\xspace}                 
 \def\POmega      {\ensuremath{\Omega}\xspace}                 
 \def\PUpsilon    {\ensuremath{\Upsilon}\xspace}
 \let\oldPi\Pi
 \def\PPi         {\ensuremath{\oldPi}\xspace}

 \def\PB      {\ensuremath{\mathrm{B}}\xspace}                 
 \def\PD      {\ensuremath{\mathrm{D}}\xspace}                 
 \def\PJ      {\ensuremath{\mathrm{J}}\xspace}                 
 \def\PK      {\ensuremath{\mathrm{K}}\xspace}                 
 \def\Pb      {\ensuremath{\mathrm{b}}\xspace}                 
 \def\Pc      {\ensuremath{\mathrm{c}}\xspace}                 
                  
 \def\Pe      {\ensuremath{\mathrm{e}}\xspace}                 
 \def\Ps      {\ensuremath{\mathrm{s}}\xspace}                 
                  
 \def\Pu      {\ensuremath{\mathrm{u}}\xspace}                 
 \def\thebaroffset{0.0em}
}
{

 \def\Peta        {\ensuremath{\eta}\xspace}

 \def\Pmu         {\ensuremath{\mu}\xspace}

 \def\Ppi         {\ensuremath{\pi}\xspace}

 \def\Ppsi        {\ensuremath{\psi}\xspace}                 
                  
 \mathchardef\PDelta="7101
 \mathchardef\PXi="7104
 \mathchardef\PLambda="7103
 \mathchardef\PSigma="7106
 \mathchardef\POmega="710A
 \mathchardef\PUpsilon="7107
 \mathchardef\PPi="7105
 \def\PB      {\ensuremath{B}\xspace}                 
 \def\PD      {\ensuremath{D}\xspace}                 
 \def\PJ      {\ensuremath{J}\xspace}                 
 \def\PK      {\ensuremath{K}\xspace}                 
 \def\Pb      {\ensuremath{b}\xspace}                 
 \def\Pc      {\ensuremath{c}\xspace}                 
                  
 \def\Pe      {\ensuremath{e}\xspace}                 
 \def\Ps      {\ensuremath{s}\xspace}                 
                  
 \def\Pu      {\ensuremath{u}\xspace}                 
 \def\thebaroffset{0.18em}
}
\newcommand{\offsetoverline}[2][\thebaroffset]{\kern #1\overline{\kern -#1 #2}}%

\makeatletter
\ifcase \@ptsize \relax
  \newcommand{\miniscule}{\@setfontsize\miniscule{4}{5}}
\or
  \newcommand{\miniscule}{\@setfontsize\miniscule{5}{6}}
\or
  \newcommand{\miniscule}{\@setfontsize\miniscule{5}{6}}
\fi
\makeatother

\DeclareRobustCommand{\optbar}[1]{\shortstack{{\miniscule (\rule[.5ex]{1.25em}{.18mm})}
  \\ [-.7ex] $#1$}}

\def\electron   {{\ensuremath{\Pe}}\xspace}

\def\epem       {{\ensuremath{\Pe^+\Pe^-}}\xspace}

\def\mup        {{\ensuremath{\Pmu^+}}\xspace}

\def\uquark    {{\ensuremath{\Pu}}\xspace}

\def\squark    {{\ensuremath{\Ps}}\xspace}

\def\cquark    {{\ensuremath{\Pc}}\xspace}

\def\bquark    {{\ensuremath{\Pb}}\xspace}

\def\pion   {{\ensuremath{\Ppi}}\xspace}
\def\piz    {{\ensuremath{\pion^0}}\xspace}
\def\pip    {{\ensuremath{\pion^+}}\xspace}
\def\pim    {{\ensuremath{\pion^-}}\xspace}

\def\kaon    {{\ensuremath{\PK}}\xspace}

\def\KorKbar {\kern \thebaroffset\optbar{\kern -\thebaroffset \PK}{}\xspace}

\def\Kp      {{\ensuremath{\kaon^+}}\xspace}
\def\Km      {{\ensuremath{\kaon^-}}\xspace}

\def\KS      {{\ensuremath{\kaon^0_{\mathrm{S}}}}\xspace}

\newcommand{\etaz}{\ensuremath{\Peta}\xspace}

\def\Dbar    {{\ensuremath{\offsetoverline{\PD}}}\xspace}
\def\D       {{\ensuremath{\PD}}\xspace}

\def\DorDbar {\kern \thebaroffset\optbar{\kern -\thebaroffset \PD}\xspace}
\def\Dz      {{\ensuremath{\D^0}}\xspace}
\def\Dzb     {{\ensuremath{\Dbar{}^0}}\xspace}
\def\Dp      {{\ensuremath{\D^+}}\xspace}
\def\Dm      {{\ensuremath{\D^-}}\xspace}

\def\DpDm    {\ensuremath{\Dp {\kern -0.16em \Dm}}\xspace}

\def\B       {{\ensuremath{\PB}}\xspace}

\def\BorBbar {\kern \thebaroffset\optbar{\kern -\thebaroffset \PB}\xspace}
\def\Bz      {{\ensuremath{\B^0}}\xspace}

\def\Bd      {{\ensuremath{\B^0}}\xspace}

\def\BdorBdbar {\kern \thebaroffset\optbar{\kern -\thebaroffset \Bd}\xspace}
\def\Bu      {{\ensuremath{\B^+}}\xspace}

\def\Bp      {{\ensuremath{\Bu}}\xspace}

\def\Bs      {{\ensuremath{\B^0_\squark}}\xspace}

\def\BsorBsbar {\kern \thebaroffset\optbar{\kern -\thebaroffset \Bs}\xspace}
\def\Bc      {{\ensuremath{\B_\cquark^+}}\xspace}

\def\jpsi     {{\ensuremath{{\PJ\mskip -3mu/\mskip -2mu\Ppsi}}}\xspace}

\def\Y#1S{\ensuremath{\PUpsilon{(#1S)}}\xspace}

\def\LorLbar     {\kern \thebaroffset\optbar{\kern -\thebaroffset \PLambda}\xspace}

\def\BF         {{\ensuremath{\mathcal{B}}}\xspace}
\def\BR         {\BF}

\newcommand{\decay}[2]{\ensuremath{\mathinner{#1\!\to #2}}\xspace}

\def\to                 {\ensuremath{\rightarrow}\xspace}

\def\Vub  {{\ensuremath{V_{\uquark\bquark}^{\phantom{\ast}}}}\xspace}
\def\Vcb  {{\ensuremath{V_{\cquark\bquark}^{\phantom{\ast}}}}\xspace}

\def\AT#1     {\ensuremath{A_{\mathrm{T}}^{#1}}\xspace}           

\def\C#1      {\ensuremath{\mathcal{C}_{#1}}\xspace}                       
\def\Cp#1     {\ensuremath{\mathcal{C}_{#1}^{'}}\xspace}                    
\def\Ceff#1   {\ensuremath{\mathcal{C}_{#1}^{\mathrm{(eff)}}}\xspace}        
\def\Cpeff#1  {\ensuremath{\mathcal{C}_{#1}^{'\mathrm{(eff)}}}\xspace}       
\def\Ope#1    {\ensuremath{\mathcal{O}_{#1}}\xspace}                       
\def\Opep#1   {\ensuremath{\mathcal{O}_{#1}^{'}}\xspace}                    

\newcommand{\aunit}[1]{\ensuremath{\text{\,#1}}}       

\newcommand{\tev}{\aunit{Te\kern -0.1em V}\xspace}
\newcommand{\gev}{\aunit{Ge\kern -0.1em V}\xspace}
\newcommand{\mev}{\aunit{Me\kern -0.1em V}\xspace}
\newcommand{\kev}{\aunit{ke\kern -0.1em V}\xspace}
\newcommand{\ev}{\aunit{e\kern -0.1em V}\xspace}
 
\newcommand{\mevc}{\ensuremath{\aunit{Me\kern -0.1em V\!/}c}\xspace}
\newcommand{\gevc}{\ensuremath{\aunit{Ge\kern -0.1em V\!/}c}\xspace}
\newcommand{\mevcc}{\ensuremath{\aunit{Me\kern -0.1em V\!/}c^2}\xspace}
\newcommand{\gevcc}{\ensuremath{\aunit{Ge\kern -0.1em V\!/}c^2}\xspace}

\def\fb   {\ensuremath{\aunit{fb}}\xspace}
\def\invfb   {\ensuremath{\fb^{-1}}\xspace}

\def\gsim{{~\raise.15em\hbox{$>$}\kern-.85em
          \lower.35em\hbox{$\sim$}~}\xspace}
\def\lsim{{~\raise.15em\hbox{$<$}\kern-.85em
          \lower.35em\hbox{$\sim$}~}\xspace}

\def\sPlot{\mbox{\em sPlot}\xspace}

\def\pt         {\ensuremath{p_{\mathrm{T}}}\xspace}

\def\evtgen     {\mbox{\textsc{EvtGen}}\xspace}

\def\tell1  {TELL1\xspace}
\def\ukl1   {UKL1\xspace}

\newcommand{\lhcborcid}[1]{\href{https://orcid.org/#1}{\hspace*{0.1em}\raisebox{-0.45ex}{\includegraphics[width=1em]{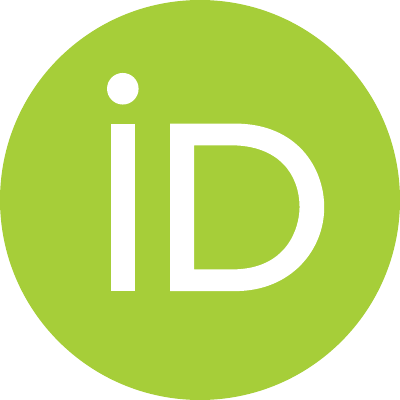}}}}

\hypersetup{
  colorlinks   = true, 
  urlcolor     = blue, 
  linkcolor    = blue, 
  citecolor    = red   
}

\ifEnableSectionTOCLinks
    \usepackage[explicit]{titlesec} 
    
    \let\oldcontentsline\contentsline
    \renewcommand

    \titleformat{\section}{\normalfont\Large\bf}{\hyperlink{tocsection.\thesection}{{\thesection} \parbox[t]{\dimexpr\textwidth-1pc}{#1}}}{1pc}{}

    \titleformat{\subsection}{\normalfont\bf}{\hyperlink{tocsubsection.\thesubsection}{{\thesubsection} \parbox[t]{\dimexpr\textwidth-1pc}{#1}}}{1pc}{}

    \titleformat{name=\section,numberless}[display]{}{}{0pt}{\normalfont\Huge\bfseries #1}
\fi

\usepackage{cite} 
\usepackage{LHCb/mciteplus}

\usepackage{longtable} 

\usepackage{xspace} 

\def\lambdab        {\ensuremath{\lambda_B}\xspace}
\def\Egamma         {\ensuremath{E^\ast_\gamma}\xspace}

\def\Bumunugamma    {\ensuremath{B^+ \to \mu^+ \nu_\mu \gamma}\xspace}
\def\Bcmunugamma    {\ensuremath{B_c^+ \to \mu^+ \nu_\mu \gamma}\xspace}
\def\Bucmunugamma    {\ensuremath{B_{(c)}^+ \to \mu^+ \nu_\mu \gamma}\xspace}
\def\munugamma      {\ensuremath{B_{(c)}^+ \to \mu^+ \nu_\mu \gamma}\xspace}
\def\ellnugamma     {\ensuremath{B^+ \to \ell^+ \nu_\ell \gamma}\xspace}

\def\sPlot          {\ensuremath{{}_{s}{\mathcal{P}lot}}\xspace}

\def\gee            {\ensuremath{\gamma_{ee}}\xspace}
\def\gcalo          {\ensuremath{\gamma_{\rm calo}}\xspace}

\def\Kstgamma       {\ensuremath{B^0 \to K^{*}(892)^0 \gamma}\xspace}

\def\wL0            {\ensuremath{\mathrm{w_{L0}}}\xspace}

\begin{document}

\renewcommand{\thefootnote}{\fnsymbol{footnote}}
\setcounter{footnote}{1}

\begin{titlepage}
\pagenumbering{roman}

\vspace*{-1.5cm}
\centerline{\large EUROPEAN ORGANIZATION FOR NUCLEAR RESEARCH (CERN)}
\vspace*{1.5cm}
\noindent
\begin{tabular*}{\linewidth}{lc@{\extracolsep{\fill}}r@{\extracolsep{0pt}}}
\ifthenelse{\boolean{pdflatex}}
{\vspace*{-1.5cm}\mbox{\!\!\!\includegraphics[width=.14\textwidth]{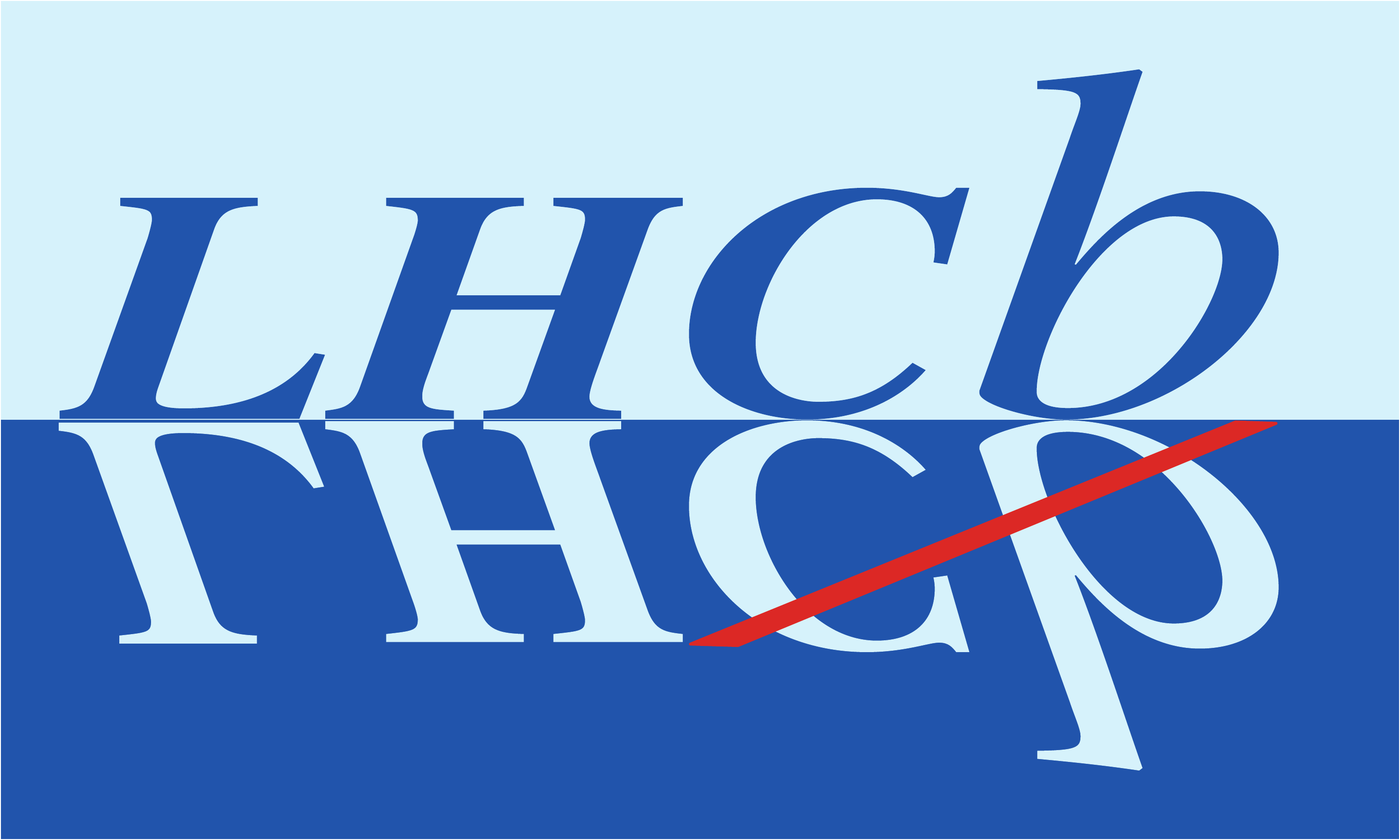}} & &}%
{\vspace*{-1.2cm}\mbox{\!\!\!\includegraphics[width=.12\textwidth]{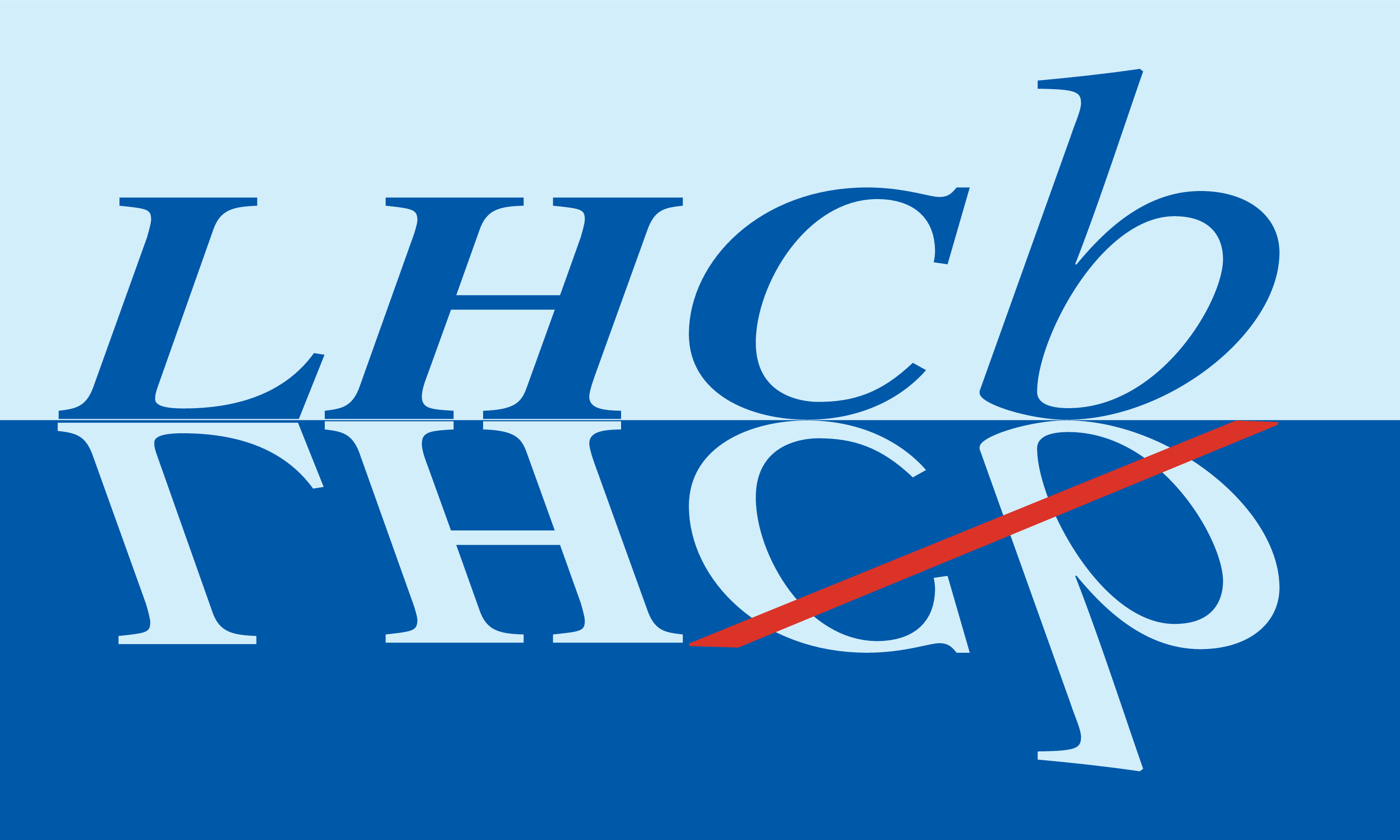}} & &}%
\\
 & & CERN-EP-2026-179 \\  
 & & LHCb-PAPER-2026-020 \\  
 & & July 14, 2026 \\ 
 & & \\
\end{tabular*}

\vspace*{4.0cm}

{\normalfont\bfseries\boldmath\huge
\begin{center}
  \papertitle 
\end{center}
}

\vspace*{2.0cm}

\begin{center}
\paperauthors\footnote{Authors are listed at the end of this letter.}
\end{center}

\vspace{\fill}

\begin{abstract}
  \noindent
  A search for the radiative leptonic decays  $B^+\to\mu^+\nu_\mu\gamma$ and 
  $B^+_c\to\mu^+\nu_\mu\gamma$ is performed using proton-proton collision data collected with the LHCb experiment at a center-of-mass energy of $13~{\rm TeV}$, corresponding to an integrated luminosity of~$5.4~{\rm fb}^{-1}$. No evidence for an excess of events over background is observed for either signal decay. Upper limits at 90\% confidence level are set on the branching fractions: 
       \begin{equation*}
        \begin{split}
            \mathcal{B}(B^+\to\mu^+\nu_\mu\gamma)_{E_\gamma^\ast > 1\,\rm{GeV}} &< 4.0 \times 10^{-6},\\
            \mathcal{B}(B_c^+\to\mu^+\nu_\mu\gamma)_{E_\gamma^\ast > 1\,\rm{GeV}} &< 1.6 \times 10^{-3},
        \end{split}
        \end{equation*}
    where the photon energy in the $B$-meson rest frame, $E_\gamma^\ast$, is required to be greater than $1~{\rm GeV}$. This constitutes the first search for these decays at a hadron collider and the first experimental investigation of the $B_c^+\to\mu^+\nu_\mu\gamma$ decay to date.
  \end{abstract}

\vspace*{2.0cm}

\begin{center}
  Submitted to
  Phys.~Rev.~Lett.
\end{center}

\vspace{\fill}

{\footnotesize 
\centerline{\copyright~\papercopyright. \href{\paperlicenceurl}{\paperlicence}.}}
\vspace*{2mm}

\end{titlepage}


\newpage
\setcounter{page}{2}
\mbox{~}


\renewcommand{\thefootnote}{\arabic{footnote}}
\setcounter{footnote}{0}


\cleardoublepage


\pagestyle{plain} 
\setcounter{page}{1}
\pagenumbering{arabic}


Precision tests of the Standard Model (SM) in the flavor sector provide highly sensitive probes for new physics, but their discovery potential is often limited by hadronic uncertainties. A crucial challenge lies in the theoretical treatment of nonperturbative quantum chromodynamics (QCD) effects in \B-meson decays, which cannot be estimated precisely. Within the widely utilized QCD factorization approach, these nonperturbative dynamics are encoded in the \B-meson light-cone distribution amplitude (LCDA)~\cite{Beneke:1999br,Korchemsky:1999qb,Lunghi:2002ju,Bosch:2003fc}. The most important \B-meson LCDA parameter for exclusive \B decay predictions is the first inverse moment of the leading-twist LCDA, \lambdab, which is related to the typical momentum of the light spectator quark in the \B meson. However, the value of \lambdab is not well known, with estimates ranging from 200 to 600\mev~\cite{Braun_2004,Wang_2015,Wang_2016,Khodjamirian_2020,Gao_2020,Janowski_2021,Mandal_2024,li2025determinationbmesondistributionamplitudes,Han:2026nvk}.

The radiative leptonic \ellnugamma decay (the inclusion of charge-conjugate processes is implied throughout this Letter) is considered the best channel to experimentally probe the \B-meson LCDA and in particular to measure \lambdab~\cite{Beneke:2011nf,Beneke:2018wjp}. 
For energetic photons, the factorization of the decay amplitude can be established, leading to its branching fraction being proportional to $1/\lambdab^2$ at leading power. However, the \ellnugamma decay remains unobserved due to its challenging final state and small predicted branching fraction of $\mathcal{O}(10^{-6})$~\cite{Beneke:2018wjp}. Previous searches have been performed by the \babar~\cite{BaBar:2009pvj} and \belle collaborations~\cite{Belle:2015mpp,Belle:2018jqd}, the latter setting the current best 90\% confidence level (CL) upper limit on the branching fraction, integrated for photon energies in the \B rest frame of $\Egamma > 1\gev$, at
$\BR(\ellnugamma)_{E_\gamma^\ast > 1\gev} < 3.0 \times 10^{-6}$ averaged over $\ell=\mu,\electron$. This upper limit is already probing the range of branching fractions expected in the SM and providing a lower bound on \lambdab at 240\mev at 90\% CL.

This Letter reports the first search for the $B^+\to\mu^+\nu_\mu\gamma$ decay at a hadron collider using the proton-proton ($pp$) collision dataset collected with the LHCb detector at a center-of-mass energy of $13\tev$ during 2016--2018, corresponding to an integrated luminosity of~$5.4 \invfb$. At the \lhcb experiment, the displaced \Bp-decay vertex cannot be reconstructed from a single charged track. The search instead relies on photons converting to an \epem pair, thus providing three tracks to identify the \Bp-decay vertex and reduce the prompt background from the $pp$ collision.
Thanks to the large boost in the forward region and the precise vertexing, LHCb can reconstruct the line of flight of the \Bp meson and compensate for the missing neutrino.
A first search for the \Bcmunugamma decay is also performed.
Although the production rate of the \Bc meson is much smaller than that of the \Bp, its radiative leptonic decay is enhanced by a factor $\vert \Vcb \vert^2 / \vert \Vub \vert^2$, where \Vcb and \Vub are elements of the Cabibbo-Kobayashi-Maskawa (CKM) matrix, and its branching fraction is predicted to be between $2\times 10^{-5}$ and $8\times 10^{-5}$~\cite{BcChiladze:1999,BcColangelog:1999,BcChang:1999,BcBarik:2008,BcWang:2015}.
Branching fractions are measured relative to that of the \Kstgamma decay, which is also reconstructed using converted photons. This decay channel is chosen for its clean experimental signature and to cancel systematic uncertainties related to the reconstruction of the photon conversions in the detector material. 

The LHCb detector is a single-arm forward spectrometer covering the pseudorapidity range $2 < \eta < 5$, described in detail in Refs.~\cite{LHCb-DP-2008-001,LHCb-DP-2014-002}. 
The online event selection is performed by a trigger~\cite{LHCb-DP-2012-004}, which consists of a hardware stage, based on information from the calorimeter and muon systems, followed by a software stage, which applies a full event reconstruction.
At the hardware trigger stage, events are required to have low detector occupancy and either a muon with high transverse momentum relative to the beamline~(\pt) or an electron with high transverse energy in the calorimeters.
The software trigger requires at least one track with large \pt and significant impact parameter with respect to any primary $pp$ collision vertex (PV). Two or more tracks are required to form a vertex that is displaced from any PV and fulfills a set of topological criteria~\cite{LHCb-PROC-2015-018}.

Simulated samples, produced with the software packages described in Refs.~\cite{Sjostrand:2007gs, Chang:2003cq, Chang:2005hq, Chang:2015qea, Wang:2012ah, Wu:2013pya,Lange:2001uf,Golonka:2005pn,Allison:2006ve,*Agostinelli:2002hh,LHCb-PROC-2011-006}, are used to model the detector response and the effect of the selection requirements.
The $B^+\to\mu^+\nu_\mu\gamma$ and $B^+_c\to\mu^+\nu_\mu\gamma$ decays are generated using the \texttt{LNuGamma} model available in \evtgen, which is based on the leading-order calculation from Ref.~\cite{Korchemsky:1999qb}. Additional simulation samples are generated for both signal decays using a uniform phase-space model to evaluate the selection efficiency in the whole phase space. The simulated samples are corrected for known differences between data and simulation in the \B-meson production kinematics and detector occupancy using data samples of $B^+\to \jpsi K^+$ and $B^+_c \to \jpsi \pi^+$ decays, following a strategy adapted from Refs.~\cite{LHCb-PAPER-2022-045,LHCb-PAPER-2022-046}. The track reconstruction efficiency and particle identification (PID) response are corrected using the methods detailed in Refs.~\cite{LHCb-DP-2013-002,LHCb-PUB-2016-021}.

Photons are reconstructed from $\gamma \to e^+e^-$ conversions in the material of the vertex detector, occurring for roughly $7\%$ of photons and denoted as \gee, with the electrons (electron refers to both electron and positron throughout this Letter) not pointing to any~PV. Due to the absence of a magnetic field, the electron tracks are largely overlapping in the vertex detector and therefore the energy deposit in each hit is doubled. Thus, the electron tracks are required to have an average energy deposit per hit at least 25\% larger than that for single tracks. Furthermore, the dielectron candidate is required to have an invariant mass below 10\mevcc, thus removing random combinations of electrons. The \Bucmunugamma candidates are reconstructed by combining the $e^+e^-$ pair with a track identified as a muon, whose \pt is larger than 1200\mevc. The reconstructed \B decay vertex is required to be significantly displaced from the associated PV, defined as that for which the difference in the vertex-fit $\chi^2$ with and without the \B candidate is the smallest. Electron and muon tracks are required to have a large impact parameter with respect to any PV and to form a vertex of good quality. To compensate for the missing momentum of the neutrino, the corrected mass, $m_{\rm corr}(\mu^+\gee)$, is used: 
\begin{equation}
\label{eq:mcorr}
    m_{\rm corr}(\mu^+\gee) \equiv \sqrt{m(\mu^+ \gamma_{ee})^2 + \vert p_\perp \vert^2} + \vert p_\perp \vert,
\end{equation} 
where $m(\mu^+ \gamma_{ee})$ is the invariant mass of the $\mu^+\gee$ system, and $\vert p_\perp \vert$ is its momentum component perpendicular to the reconstructed \B-meson flight direction. To improve the separation in $m_{\rm corr}(\mu^+\gee)$ of \Bucmunugamma candidates from the much more abundant $B^+\to \mu^+\nu_\mu\piz$ background candidates with $\piz \to \gee\gamma$, possible energy loss of the electrons through bremsstrahlung is not recovered, as the other photon from the \piz decay could be wrongly identified as a bremsstrahlung photon. 
An approximation of the \B-candidate momentum vector in the laboratory frame is obtained as in Ref.~\cite{LHCb-PAPER-2015-025} and is used to calculate the photon energy in the \B rest frame, \Egamma, which is required to be larger than~1\gev.

To suppress specific types of background processes, three multivariate classifiers are used. A first classifier based on the \texttt{XGBoost} algorithm~\cite{XGBoost} is used to separate ${\Bucmunugamma}$ signal candidates from combinatorial background based on the decay topology, kinematics of the muon candidates, and the fit quality of the $\mu^+\gee$ vertex. Simulated \Bumunugamma decays are used as signal proxy, and candidates with $m_{\rm corr}(\mu^+\gee)$ greater than 6500\mevcc in data passing all other selection requirements are used as background.
A second classifier, also based on the \texttt{XGBoost} algorithm, is trained to suppress background where the converted photon originates from a neutral pion decay. Additional photon candidates reconstructed in the calorimeter, \gcalo, are searched in a cone close to the converted photon candidate and are used to form $\piz \to\gee \gcalo$ candidates. The classifier is trained to identify and suppress decays of neutral pions based on the invariant mass and momentum of \piz candidates, the number of reconstructed photons in the cone and the \pt asymmetry between the two photons. Simulated samples of \Bumunugamma and inclusive charmless semileptonic $B$-meson decays are used as signal and background proxies, respectively.
The third classifier is used to suppress partially reconstructed background candidates with additional charged tracks as, for example, from $\Bp \to \Dzb \mup \nu_\mu$ with \decay{\Dzb}{\Kp \pim \piz}. While there is substantial overlap with the other background categories, this dedicated classifier improves the background suppression. A Gradient Boosted Decision tree is trained to determine if the signal tracks are isolated from all remaining tracks in the rest of the event, following the approach developed by previous \lhcb analyses \cite{LHCb-PAPER-2015-025, LHCb-PAPER-2020-038}. To avoid bias from overtraining, a $k$-folding approach~\cite{10.1145/307400.307439} with $k=5$ is used for all three classifiers. The selection on the classifier responses is optimized such that it minimizes the expected upper limits on the signal branching fractions. This selection reduces the background by about 98\% while retaining 22\% of \Bumunugamma and 8\% of \Bcmunugamma decays.
To avoid experimenter’s bias, the region $4500\mevcc < m_{\rm corr}(\mu^+\gee) < 6500\mevcc$ was not examined until the full analysis procedure had been finalized.

Candidates of the normalization channel \Kstgamma are reconstructed from $\gamma\to e^+e^-$ conversions in the material of the vertex detector with the selection on the electrons fully aligned with the \munugamma signal selection. The $K^{*}(892)^0$ candidate is reconstructed as $K^+\pi^-$ with an invariant mass within 100\mevcc of its known mass~\cite{PDG2024}. The rest of the selection follows Ref.~\cite{LHCb-PAPER-2020-020}. 

Selected signal candidates in the range ${m_{\rm corr}(\mu^+\gee) \in [4000,7500]\mevcc}$ are retained and the yields of \Bumunugamma and \Bcmunugamma signal decays are obtained simultaneously by fitting a sum of histogram templates binned in $m_{\rm corr}(\mu^+\gee)$ to the data. The remaining background is dominated by candidates in which the converted photon does not directly originate from the decay of a \B meson but rather from the decay of a \piz or \etaz meson, hereafter referred to as \piz and \etaz background, respectively. Besides backgrounds from {CKM-suppressed} $\bar b\to \bar u\mu^+\nu_\mu$ transitions, those from CKM-favored $\bar b\to \bar c\mu^+\nu_\mu$ transitions, such as $B^+\to  \Dzb(\KS\pi^0)\mu^+\nu_\mu$ decays, also contribute, as well as random combinations of a $\pi^0$ or $\eta$ meson with a muon. 

\begin{figure}[tb]
    \centering
    \includegraphics[height=5cm]{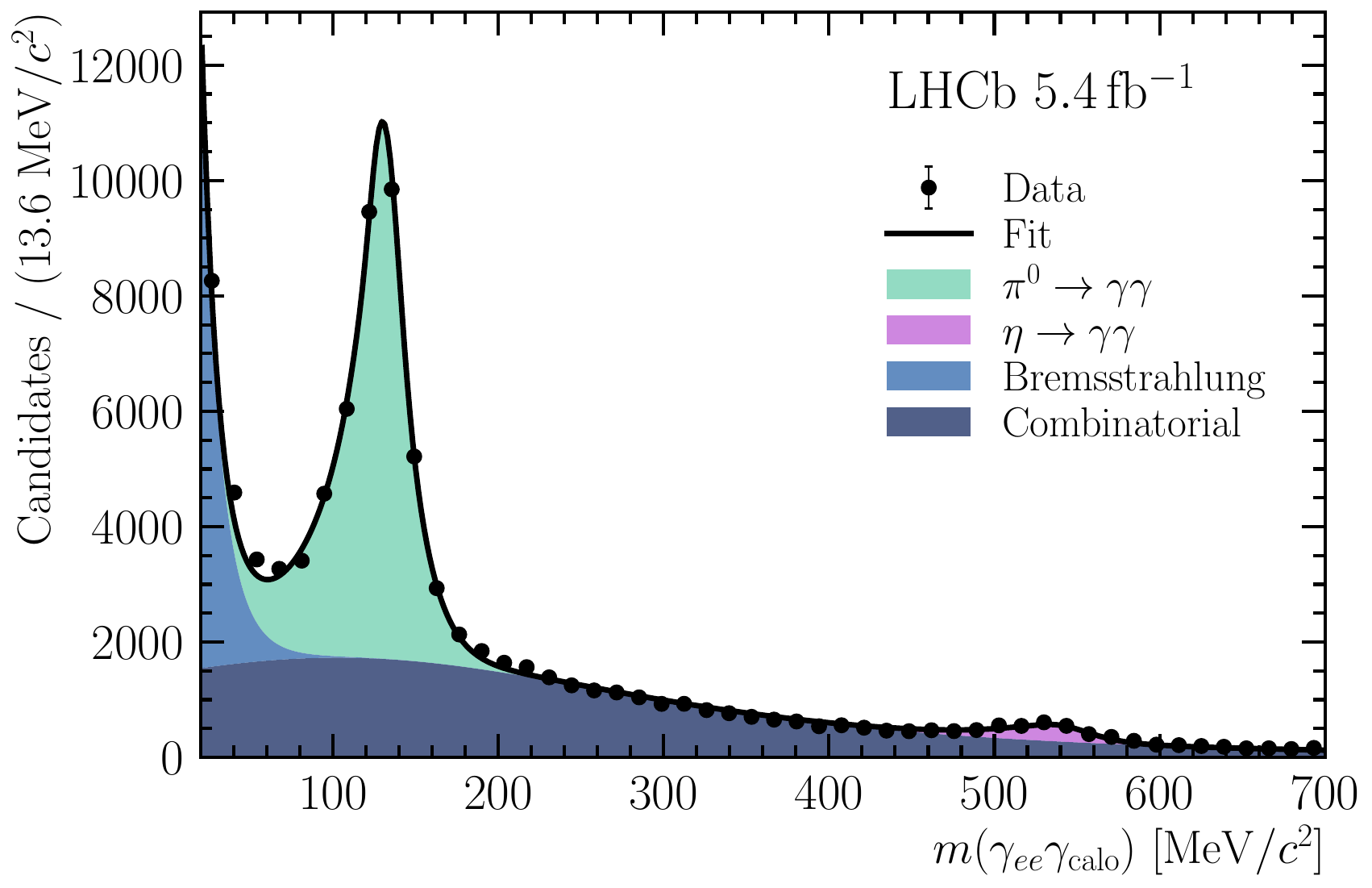}
     \includegraphics[height=5cm]{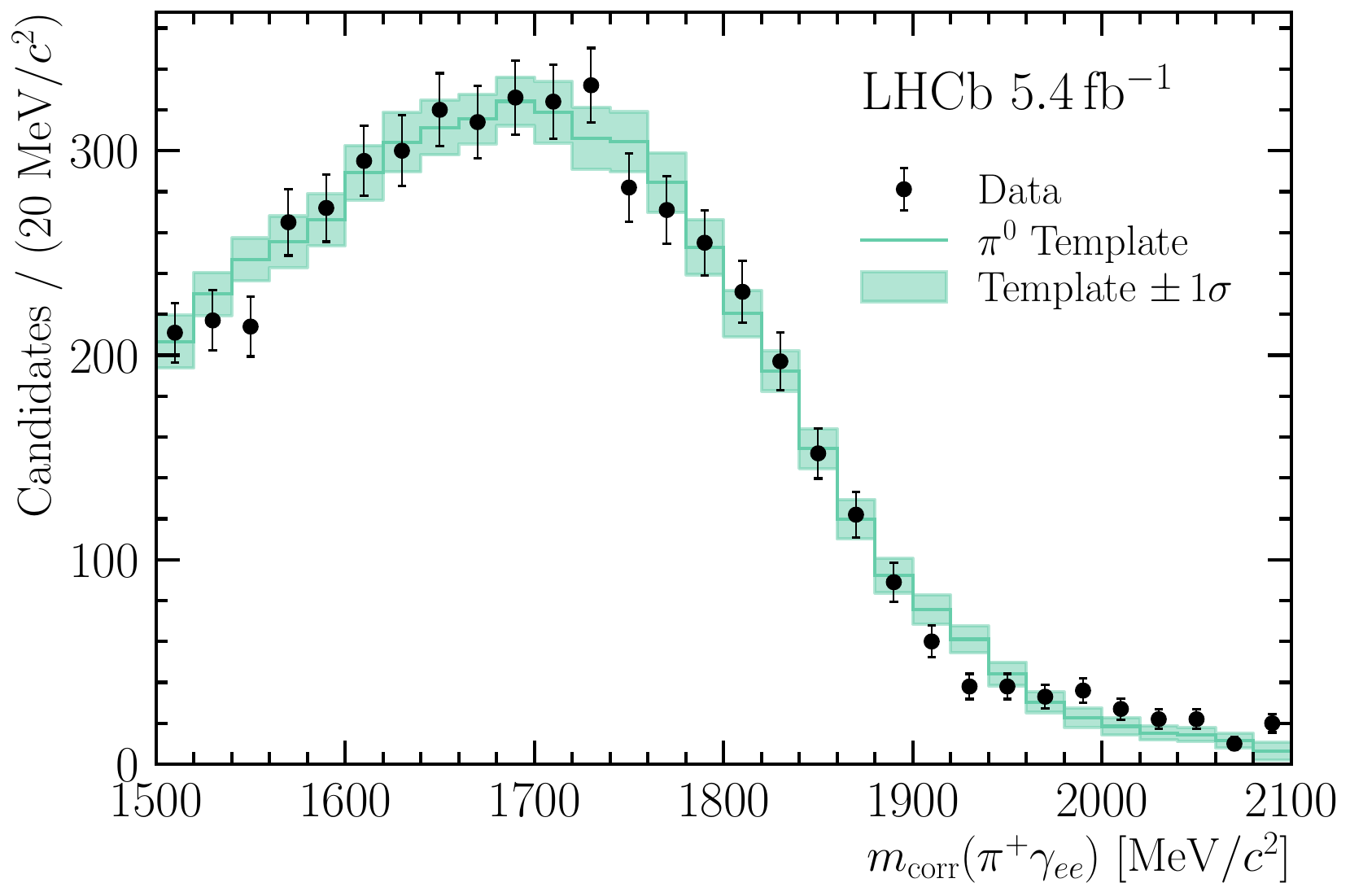}
    \caption{(Left) Distribution of $m(\gamma_{ee}\gamma_{\rm calo})$ for $B^+\to\mu^+\nu_\mu\gamma_{ee}\gamma_{\rm calo}$ candidates in data together with the result of the fit. 
    (Right) Distribution of $m_{\rm corr}(\pi^+\gamma_{ee})$ in $D^0\to K^-\pi^+\pi^0$ decays reconstructed as $D^0\to K^-\pi^+\gamma_{ee}$ as observed (black) and predicted (green).
    }
    \label{fig:piz_bkgd}
\end{figure}

The background templates for fitting the $m_{\rm corr}(\mu^+\gee)$ distribution are obtained from data to include all sources of \piz and \etaz background. For this, the $\mu^+\pi^0$ and $\mu^+\eta$ candidates are explicitly reconstructed in the dataset of signal candidates, combining the converted photon \gee with an additional \gcalo photon. 
An unbinned maximum-likelihood fit to the diphoton invariant-mass distribution, $m(\gee\gcalo)$, is performed. The fit model includes four components: the two neutral mesons \piz and \etaz; the component describing the case where the \gcalo is a bremsstrahlung photon emitted by one of the two electrons; and the combinatorial background. The shapes are obtained from simulation. The fit is performed in four bins of $\gee\gcalo$ momentum, and the combined result is shown in Fig.~\ref{fig:piz_bkgd}~(left). The \sPlot technique~\cite{Pivk:2004ty} is then used to obtain the background-subtracted $m_{\rm corr}(\mu^+\gee)$ distributions of $\mu^+\piz$ and $\mu^+\etaz$ systems.
The efficiency of reconstructing the additional \gcalo photon, and hence the $\piz\to\gee\gcalo$ and $\etaz\to\gee\gcalo$ decays, is correlated to the fraction of the \piz (\etaz) momentum carried by the converted photon. It therefore varies across the $m_{\rm corr}(\mu^+\gee)$ distribution, and efficiency corrections detailed in the End Matter are applied separately for the \piz and \etaz backgrounds.
The resulting samples of background-subtracted and efficiency-corrected $\mu^+\pi^0$ and $\mu^+\eta$ candidates are used to model the $m_{\rm corr}(\mu^+\gee)$ distribution of all remaining \piz and \etaz backgrounds. The $m_{\rm corr}(\mu^+\gee)$ templates are smoothed by Gaussian convolution with a 100\mevcc width to reduce bin-by-bin statistical fluctuations without distorting physical features in the distribution. The statistical uncertainties on the templates are estimated using a bootstrapping method~\cite{efron:1979}, resampling the data~100 times. After taking into account the efficiency, the number of \piz and \etaz backgrounds in data are expected to be $1400\pm70$ and~$140\pm40$, respectively.

The procedure for obtaining the $m_{\rm corr}(\mu^+\gamma_{ee})$ distribution and level of \piz background is validated using a data sample of $D^0\to K^-\pi^+\pi^0$ decays, reconstructed as $D^0 \to K^-\pi^+\gee$ candidates, where the \Dz originates from a $D^{\ast+}\to D^0 \pi^+$ decay. This \Dz decay is particularly well suited for the check due to its large yield and the fact that, after the selection process, the momentum distribution of the \gee is similar to that of the signal. Moreover, the contamination from genuine $D^0\to K^-\pi^+\gamma$ decays can safely be neglected since its branching fraction is about two orders of magnitude smaller than for $D^0 \to K^-\pi^+\pi^0$~\cite{PDG2024}. 
In order to mimic the signal analysis, the corrected mass for a missing kaon, $m_{\rm corr}(\pi^+\gamma_{ee})$, is constructed for the decay $D^0 \to K^-\pi^+\pi^0$. The observed $m_{\rm corr}(\pi^+\gamma_{ee})$ distribution and the derived template for $D^0 \to K^-\pi^+\pi^0$ decays are shown in Fig.~\ref{fig:piz_bkgd}~(right). They indicate excellent agreement with a $\chi^2/$ndf value of 1.07, corresponding to a $p$-value of 36\%.

The subdominant background originating from the misidentification of the muon candidate, for example from $B^+\to \KS\pi^+\gee$ decays, is modeled using the data. Data samples enriched in misidentified hadrons are obtained by inverting the PID selection requirement on the muon candidate and further divided into $\pi^+$, $K^+$ and $p$ species. Each of the candidates is weighted by the probability of misidentifying a hadron of the respective species as a muon, which is determined from control samples~\cite{LHCb-PUB-2016-021}. With this method, both the $m_{\rm corr}(\mu^+\gee)$ shape and the yield of the misidentification background are predicted. Combining all sources of hadron-to-muon misidentification, $10.1\pm1.0$ background events are expected. As the shape of the misidentification background overlaps with the \Bcmunugamma signal shape, this background component is included in the fit despite its small total yield. Other backgrounds, such as $B^+_{(c)} \to \tau^+ \nu_\tau \gamma$ with $\tau^+ \to \mu^+ \nu_\mu \bar\nu_\tau$, CKM-favored $\bar b\to \bar c\mup\nu_\mu$ decays with radiative charm-meson decays or the misidentification of electron candidates, are estimated to be negligible. 

The extended binned maximum-likelihood fit of the $m_{\rm corr}(\mu^+\gamma_{ee})$ distribution includes the two signal components, the \piz and \etaz backgrounds, and the muon misidentification background. The \piz and \etaz background yields are allowed to vary, with a Gaussian constraint applied on their relative fraction based on the abovementioned prediction. The muon misidentification yield is also constrained to the value and uncertainty determined from the control samples with a Gaussian. 
The $m_{\rm corr}(\mu^+\gee)$ templates for the signal components of \Bumunugamma and \Bcmunugamma decays are derived from simulation. Template fluctuations due to the limited size of the \Bcmunugamma simulation samples are smoothed by convolution with a Gaussian of width 100\mevcc. 
The statistical template uncertainties are accounted for by introducing additional nuisance parameters as proposed in Ref.~\cite{Conway:2011in} and included in the statistical uncertainty.
Potential biases in the fit procedure are evaluated through pseudoexperiments with different signal strengths injected, ranging from a background-only hypothesis to a signal strength that would lead to an observation with a significance of approximately five standard deviations. The fitted \Bumunugamma and \Bcmunugamma yields exhibit small negative biases of $-8.1 \pm 0.9$ and $-1.0\pm0.3$, respectively. These biases are independent of the injected signal strength and originate from the uncertainties in the \piz background template within the \Bucmunugamma signal region.
To correct for this bias, for each mode, an upward shift on the fitted yield by the size of the bias is applied a posteriori, and the uncertainty of the bias is assigned as a systematic uncertainty. 

\begin{figure}[tb]
    \centering
    \includegraphics[height=5cm]{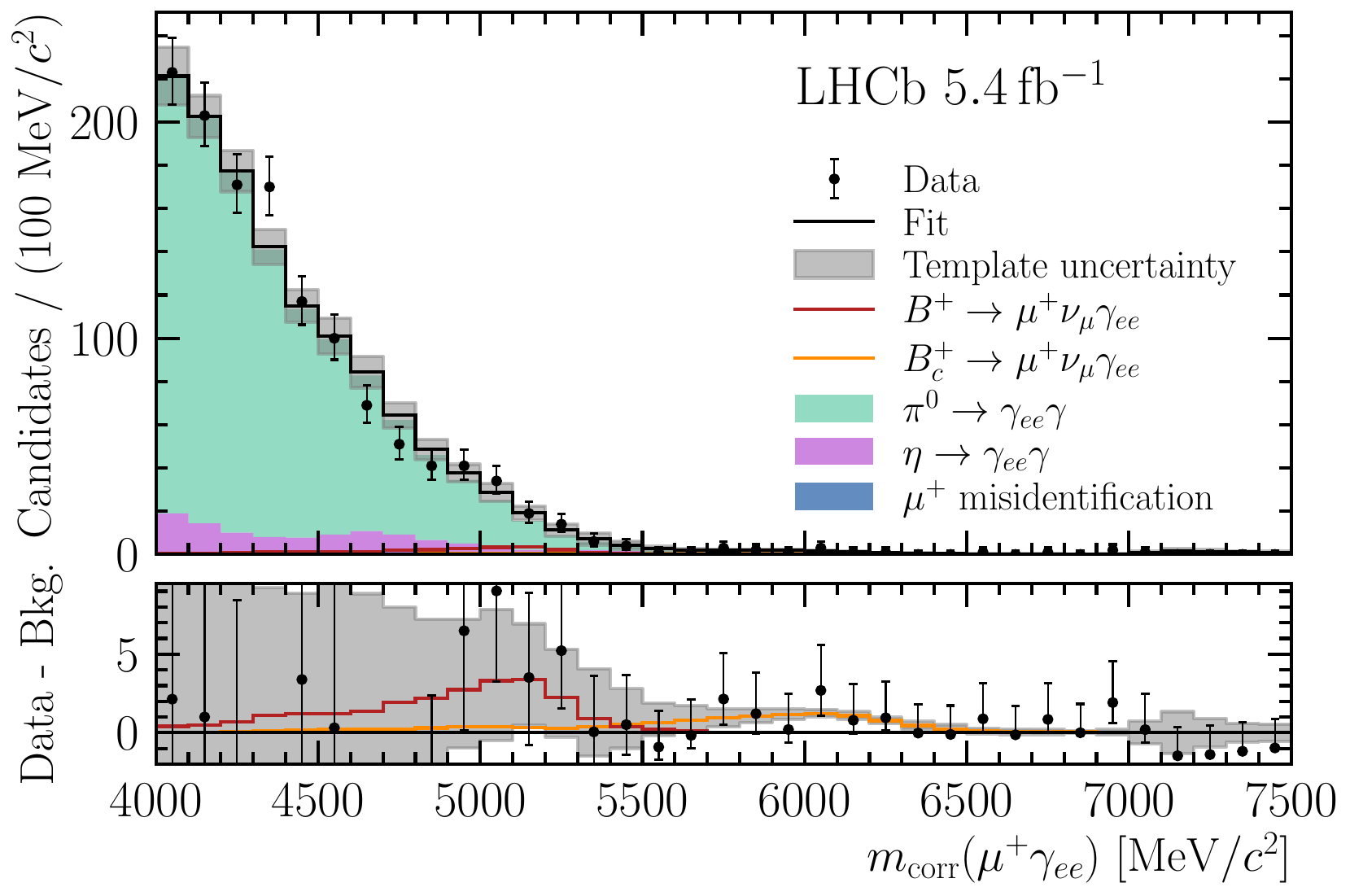}
     \includegraphics[height=5cm]{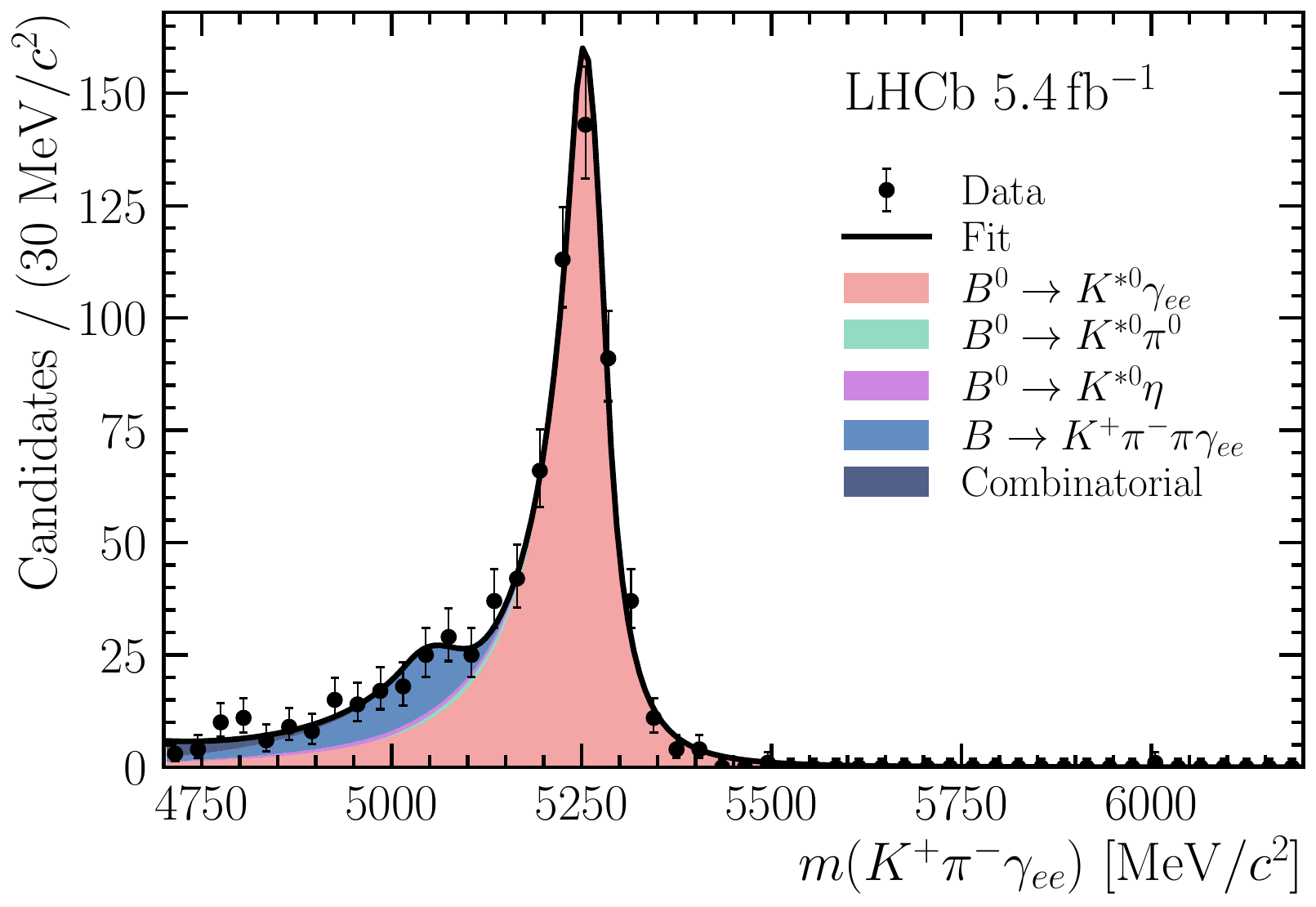}
    \caption{(Left) Corrected-mass distribution of $\mu^+\gamma_{ee}$ candidates with the fit components overlaid. The difference between data and the background component is shown below. (Right) Invariant-mass distribution of $\Kp \pim \gamma_{ee}$ candidates with the fit components overlaid.}
    \label{fig:massFits}
\end{figure}

Figure~\ref{fig:massFits}~(left) shows the $m_{\rm corr}(\mu^+\gee)$ distribution and the result of the fit overlaid. The  \Bumunugamma and \Bcmunugamma yields, after the bias corrections, are $ 32 \pm 24$ and~$13^{+6}_{-5}$, respectively. The two yields exhibit a small anticorrelation of 10\%. The \piz and \etaz background yields are respectively $1130 \pm 50$ and $106 \pm 29$. They differ from the expectations because of different kinematics of the \piz and \etaz mesons in data and simulation, the latter using exclusively the decays $\Bp \to \mup\nu_\mu\piz$ and $\Bp\to \mup\nu_\mu\etaz$, respectively.
The fit of the reconstructed invariant mass of the normalization channel follows the strategy in Ref.~\cite{LHCb-PAPER-2020-020} and is shown in Fig.~\ref{fig:massFits} (right), with a yield of $616 \pm 29$ \Kstgamma decays. 
The branching fractions of the \Bumunugamma and \Bcmunugamma signal decays are measured with respect to the \Kstgamma decay as
\begin{eqnarray}
   \BR(\Bucmunugamma) &=& \frac{N_{\Bucmunugamma}}{N_{\Kstgamma}} \times \frac{\varepsilon_{\Kstgamma}}{\varepsilon_{\Bucmunugamma}} \times \frac{f_{d}}{f_{u(c)}} \\&\times& \BR\left({\Kstgamma}\right)\times \BR\left({K^{*}(892)^0\to \Kp \pi^-}\right). \nonumber
   \label{eq:fromNtoBR} 
\end{eqnarray}
The measured branching fractions depend on the fragmentation fractions $f_d$, $f_u$ and $f_c$, which are the probabilities for a $b$ quark to hadronize into a \Bz, \Bp or \Bc meson. The ratio $f_c/(f_u+f_d) = (3.78\pm0.90)\times 10^{-3}$ has been determined in Ref.~\cite{LHCb-PAPER-2019-033} and $f_u=f_d$, assuming isospin symmetry.
The \Kstgamma branching fraction is taken from Ref.~\cite{PDG2024}.
The ratios of efficiencies $\varepsilon_{\Kstgamma} / \varepsilon_{\Bumunugamma}$ and $\varepsilon_{\Kstgamma} / \varepsilon_{\Bcmunugamma}$ are determined using simulation to be $1.32 \pm 0.05$ and $12.7 \pm 0.9$, respectively, where the uncertainties are statistical only. 
The efficiency of selecting \Bcmunugamma decays is about an order of magnitude smaller than for the \Bu decay due to the significantly shorter lifetime of the \Bc meson.
These efficiencies are valid for the \munugamma decay described by the \texttt{LNuGamma} model based on Ref.~\cite{Korchemsky:1999qb}, but the result of this analysis can be reinterpreted for any model using the efficiency maps provided in the End Matter.

\begin{table}[tb]
    \centering
    \footnotesize
    \caption{Summary of the systematic and external uncertainties. }
    \begin{tabular}{ccc}
    \hline
    Source & $\mathcal{B}(\Bu)$ [$10^{-6}$] & $\mathcal{B}(\Bc)$ [$10^{-4}$]\\
    \hline
    \piz / \etaz background model & 0.76 & 1.01\\
    Signal model & 0.21 & 0.92\\
    Fit bias & 0.05 & 0.23 \\
    Muon misidentification & 0.07 & 0.63 \\
    Efficiency calibration & 0.06 & 0.32\\
Selection efficiency & 0.02 & 0.10\\
Simulation sample size & 0.07 & 0.74\\
Normalization yield & 0.09 & 0.46 \\
    \hline
$\mathcal{B}(\Kstgamma)$~\cite{PDG2024} & 0.11 & 0.59\\
    $f_d/f_c$~\cite{LHCb-PAPER-2019-033} & $-$ & 2.33 \\
    \hline
    Total systematic & 0.80 & 1.79\\
    Total external & 0.11 & 2.40\\
    \hline
    \end{tabular}
    \label{tab:systematics}
\end{table}

A summary of the systematic uncertainties is given in Table~\ref{tab:systematics}, yielding a total uncertainty of $0.8\times 10^{-6}$ on the \Bumunugamma and $3.0\times 10^{-4}$ on the \Bcmunugamma branching fractions. 
The largest systematic uncertainty arises from the knowledge of the $m_{\rm corr}(\mu^+\gee)$ model for \piz and \etaz backgrounds, which is evaluated in two steps. First, an alternative efficiency correction for reconstructing the \gcalo photon from $\piz\to\gee\gcalo$ and $\etaz\to\gee\gcalo$ decays is applied. Secondly, the background templates are smoothed with a Gaussian function with the width reduced to 50\mevcc. Further details are provided in the End Matter.
The second largest systematic uncertainty originates from the uncertainty on the \B-meson LCDA, which can lead to a significant change in the decay kinematics.
This induces a model dependence on the signal efficiency as well as the $m_{\rm corr}(\mu^+\gee)$ shape. The effect on the efficiency can be corrected using the information provided in the End Matter, whereas the systematic uncertainty due to the signal shape variation is assessed from the largest effect observed when using signal templates weighted to match the calculation in Ref.~\cite{Beneke:2018wjp} for values of $\lambda_B$ between 200 and 600\mev.
For the model of \Bcmunugamma decays, an additional uncertainty related to the smoothing procedure of the $m_{\rm corr}(\mu^+\gee)$ template is assigned.
The uncertainty on the observed fit bias is assigned as a systematic uncertainty, and the uncertainty related to the small number of expected muon misidentification background is obtained from doubling the expected background yield.

The finite size of the data control samples used to calibrate the track reconstruction, trigger and PID efficiencies introduces a systematic uncertainty, which, due to differences in the selection and reconstruction of the signal and normalization decays, does not fully cancel. It is propagated to the result on the ratio of branching fractions using a bootstrapping method~\cite{efron:1979}. The modeling of nuclear interactions in the simulation introduces an additional uncertainty on the efficiency calibration, which is accounted for by using the procedure outlined in Ref.~\cite{LHCb-DP-2013-002}. Additional systematic uncertainties account for residual differences in the selection of candidates between the signal and normalization channels, the finite size of the simulation samples as well as the uncertainty on the normalization yield.
The external uncertainties comprise the uncertainties on the \Kstgamma branching fraction from Ref.~\cite{PDG2024} and the ratio of fragmentation fractions $f_d/f_c$ from Ref.~\cite{LHCb-PAPER-2019-033}.

The branching fractions are determined to be
\begin{equation*}
\begin{split}
        \mathcal{B}(\Bumunugamma)_{E_\gamma^\ast > 1\,\rm{GeV}} &= (1.9 \pm 1.5 \pm 0.8 \pm 0.1)\times 10^{-6},\\
        \mathcal{B}(\Bcmunugamma)_{E_\gamma^\ast > 1\,\rm{GeV}} &= (9.8 {}^{+4.6}_{-3.8} \pm 1.8\pm 2.4)\times 10^{-4},
    \end{split}
\end{equation*}
where the first uncertainty is statistical, the second is systematic, and the third is due to external uncertainties.
The log-likelihood profiles as functions of the branching fractions are shown in Fig.~\ref{fig:nll} of the End Matter. These branching fractions are compatible with zero within 1.3 and 2.4 standard deviations for the \Bu and \Bc decays, respectively. 
Consequently, upper limits on the \Bu and \Bc branching fractions are evaluated using the CL$_s$ method~\cite{CLs} with a one-sided test statistic~\cite{Cowan:2011}. The CL$_s$ curves are shown in Fig.~\ref{fig:limits} of the End Matter from which the limits on the \Bumunugamma and \Bcmunugamma branching fractions are found to be
\begin{equation*}
 \begin{split}
            \mathcal{B}(\Bumunugamma)_{E_\gamma^\ast > 1\,\rm{GeV}} &< 4.0\,(4.6)\times 10^{-6},\\
            \mathcal{B}(\Bcmunugamma)_{E_\gamma^\ast > 1\,\rm{GeV}} &< 1.6\,(1.9)\times 10^{-3},
        \end{split}
\end{equation*}
at 90\% (95\%) CL. The sensitivity on the branching fraction of the ${\Bumunugamma}$ decay is comparable with the result reported in Ref.~\cite{Belle:2018jqd}. For the \Bcmunugamma branching fraction, an upper limit is determined for the first time. These searches are performed for the first time at a hadron collider and are expected to benefit from the increased luminosity and improved performance of the upgraded LHCb detector in Run 3~\cite{LHCb-DP-2022-002}.


\section*{Acknowledgements}
%
%
\noindent We express our gratitude to our colleagues in the CERN
accelerator departments for the excellent performance of the LHC. We
thank the technical and administrative staff at the LHCb
institutes.
We acknowledge support from CERN and from the national agencies:
ARC (Australia);
CAPES, CNPq, FAPERJ and FINEP (Brazil); 
MOST and NSFC (China); 
CNRS/IN2P3 and CEA (France);  
BMFTR, DFG and MPG (Germany);
NKFIH (Hungary);              
INFN (Italy); 
NWO (Netherlands); 
MNiSW and NCN (Poland); 
MEC/IFA (Romania); 
MICIU and AEI (Spain);
SNSF and SER (Switzerland); 
NASU (Ukraine); 
STFC (United Kingdom); 
DOE NP and NSF (USA).
We acknowledge the computing resources that are provided by ARDC (Australia), 
CBPF (Brazil),
CERN, 
IHEP and LZU (China),
IN2P3 (France), 
KIT and DESY (Germany), 
INFN (Italy), 
SURF (Netherlands),
Polish WLCG (Poland),
IFIN-HH (Romania), 
PIC (Spain), CSCS (Switzerland), 
GridPP (United Kingdom),
and NSF (USA).  
We are indebted to the communities behind the multiple open-source
software packages on which we depend.
Individual groups or members have received support from
RTP (Australia), 
FWO Odysseus grant G0ASD25N (Belgium), 
Key Research Program of Frontier Sciences of CAS, CAS PIFI, CAS CCEPP (China); 
Minciencias (Colombia);
EPLANET, Marie Sk\l{}odowska-Curie Actions, ERC and NextGenerationEU (European Union);
A*MIDEX, ANR, IPhU and Labex P2IO, and R\'{e}gion Auvergne-Rh\^{o}ne-Alpes (France);
Alexander-von-Humboldt Foundation (Germany);
ICSC (Italy); 
Severo Ochoa and Mar\'ia de Maeztu Units of Excellence, GVA, XuntaGal, GENCAT, InTalent-Inditex and Prog.~Atracci\'on Talento CM (Spain);
the Leverhulme Trust, the Royal Society and UKRI (United Kingdom).

\newpage 
\section*{End Matter}

\appendix
\subsection*{Additional information on the \piz and \etaz background templates}
From the fit of the diphoton invariant-mass distribution shown in Fig.~\ref{fig:piz_bkgd} (left), total yields of $33900\pm300$ $\mup\piz$ and $1740\pm80$ $\mup\etaz$ decays are found. As no correlation between $m(\gee\gcalo)$ and $m_{\rm corr}(\mup\gee)$ is observed, the \sPlot technique can be applied to obtain the $m_{\rm corr}(\mup\gee)$ distributions for background-subtracted $\mup\piz$ and $\mup\etaz$ samples. The efficiencies of reconstructing the \gcalo photon and hence the $\piz\to\gee\gcalo$ and $\etaz\to\gee\gcalo$ decays are obtained as a function of $\pt(\gcalo)$, from the simulation of $\Bp\to \mup\nu_\mu\piz$ and $\Bp\to \mup\nu_\mu\etaz$ decays, respectively. The inverse of this efficiency is applied as a per-event weight to the background-subtracted samples of $\mup\piz$ and $\mup\eta$ combinations. Candidates for which the additional photon has too small $\pt(\gcalo)$ to be reconstructed are accounted for using an extrapolation of the \gcalo kinematics established in $\Bp\to \mup\nu_\mu\piz$ and $\Bp\to \mup\nu_\mu\etaz$ simulation. The full procedure is validated by applying the same efficiency corrections and extrapolations to the data of the control channel $D^{\ast +}\to \Dz(\Km\pip\piz)\pip$, as shown in Fig.~\ref{fig:piz_bkgd} (right), with the only modification of reducing the width of the Gaussian function to smooth the template to 20\mevcc.

The systematic uncertainty associated with the \piz and \etaz templates is evaluated in two steps. A first set of alternative background templates is generated from extending the extrapolation of the \gcalo kinematics across a much wider kinematic range. A second set of alternative background templates is generated by reducing the width of the Gaussian function, used to reduce statistical bin-by-bin fluctuations, to 50\mevcc. In total, 1000 pseudoexperiments are conducted, with pseudodata generated from a linear interpolation between the nominal and alternative background templates with a flat prior. The quadratic sum of the shift in the central values and the change in the spread of the signal branching fractions is assigned as a systematic uncertainty.

\subsection*{Additional information on the fit result and limit setting} 
\begin{figure}[h]
   \centering
    \includegraphics[width=0.4\textwidth]{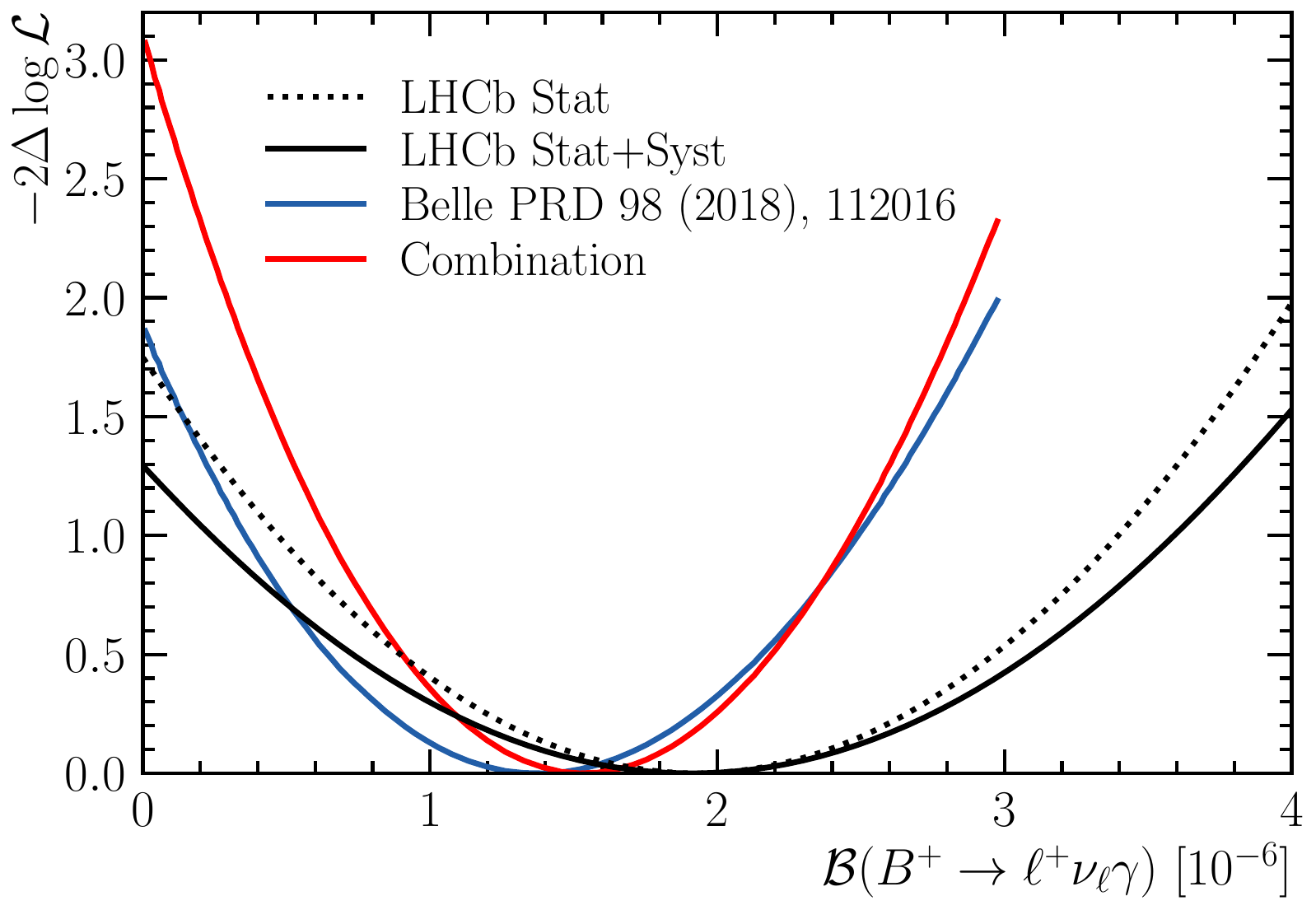}
    \includegraphics[width=0.4\textwidth]{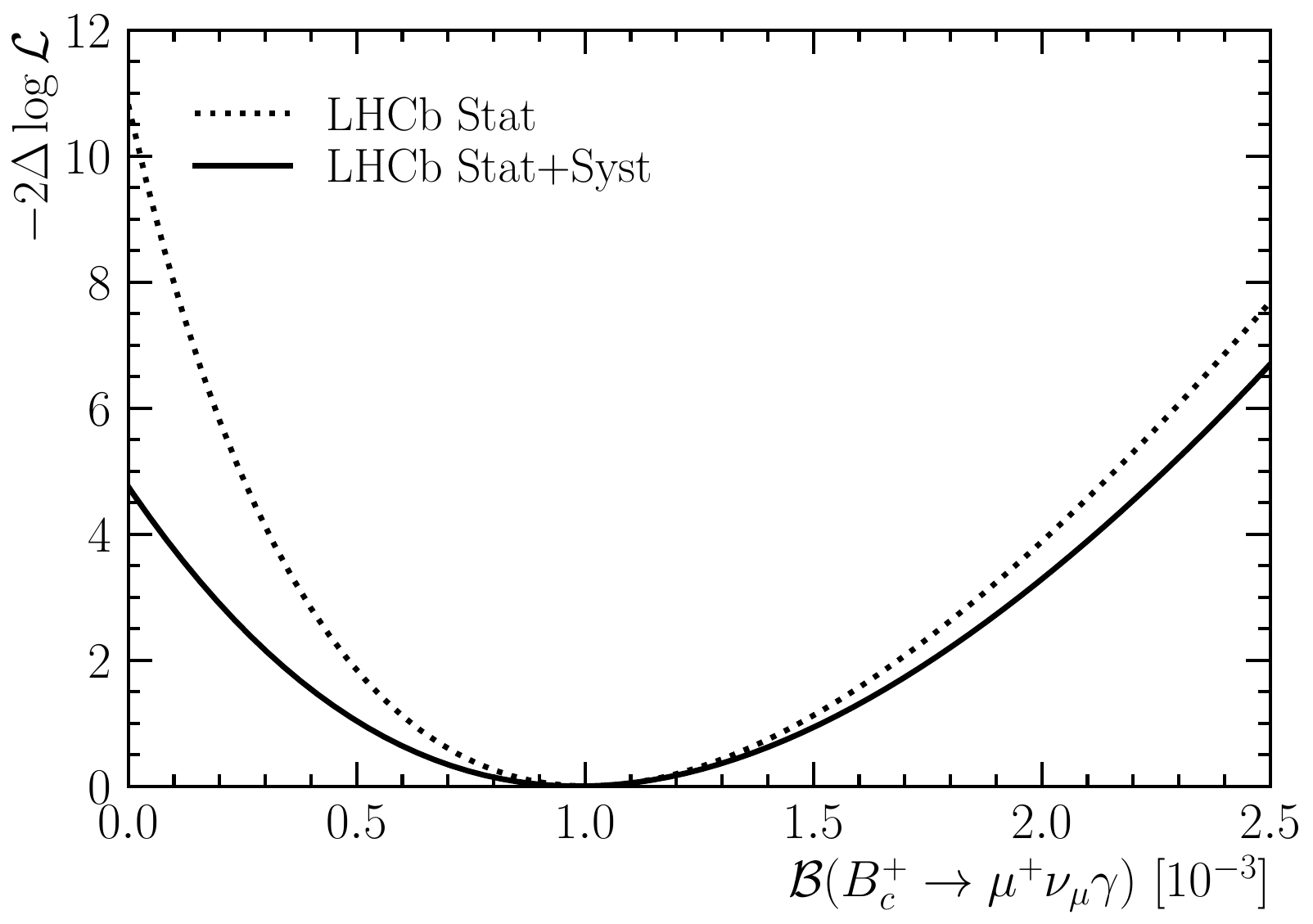}\\
    \caption{Profiles of $-2\Delta\mathcal{\log L}$ for (left) $B^+\to\ell^+\nu_\ell\gamma$ and (right) $B_c^+\to\mu^+\nu_\mu\gamma$ branching fractions (solid) with and (dashed) without systematic variations included. The profile likelihood for the $B^+\to\ell^+\nu_\ell\gamma$ branching fraction in blue is taken from Ref.\cite{Belle:2018jqd} and the combination is shown in red, assuming lepton universality.}
    \label{fig:nll}
\end{figure}

\begin{figure}[h]
    \centering
    \includegraphics[height=5cm]{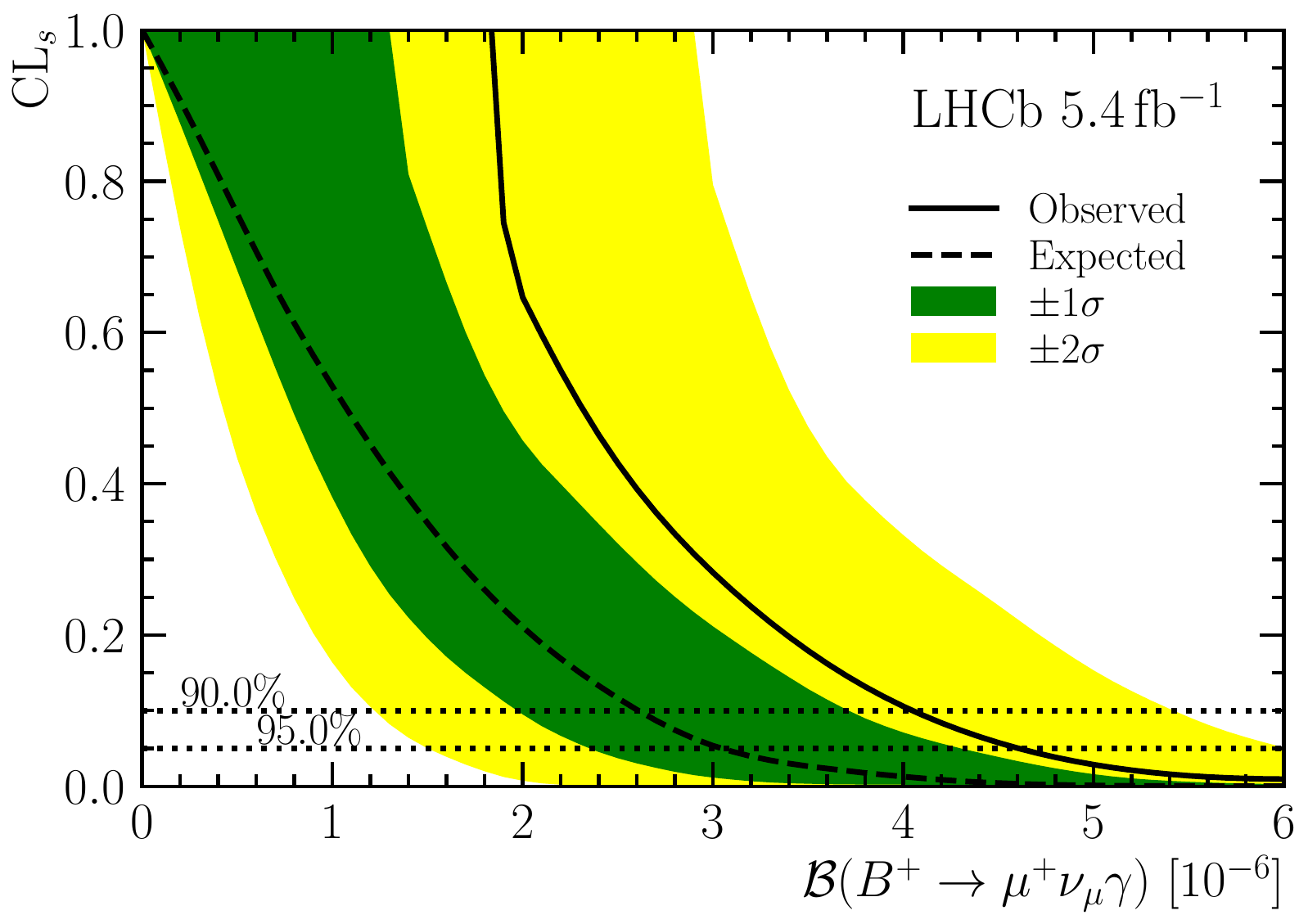}
    \includegraphics[height=5cm]{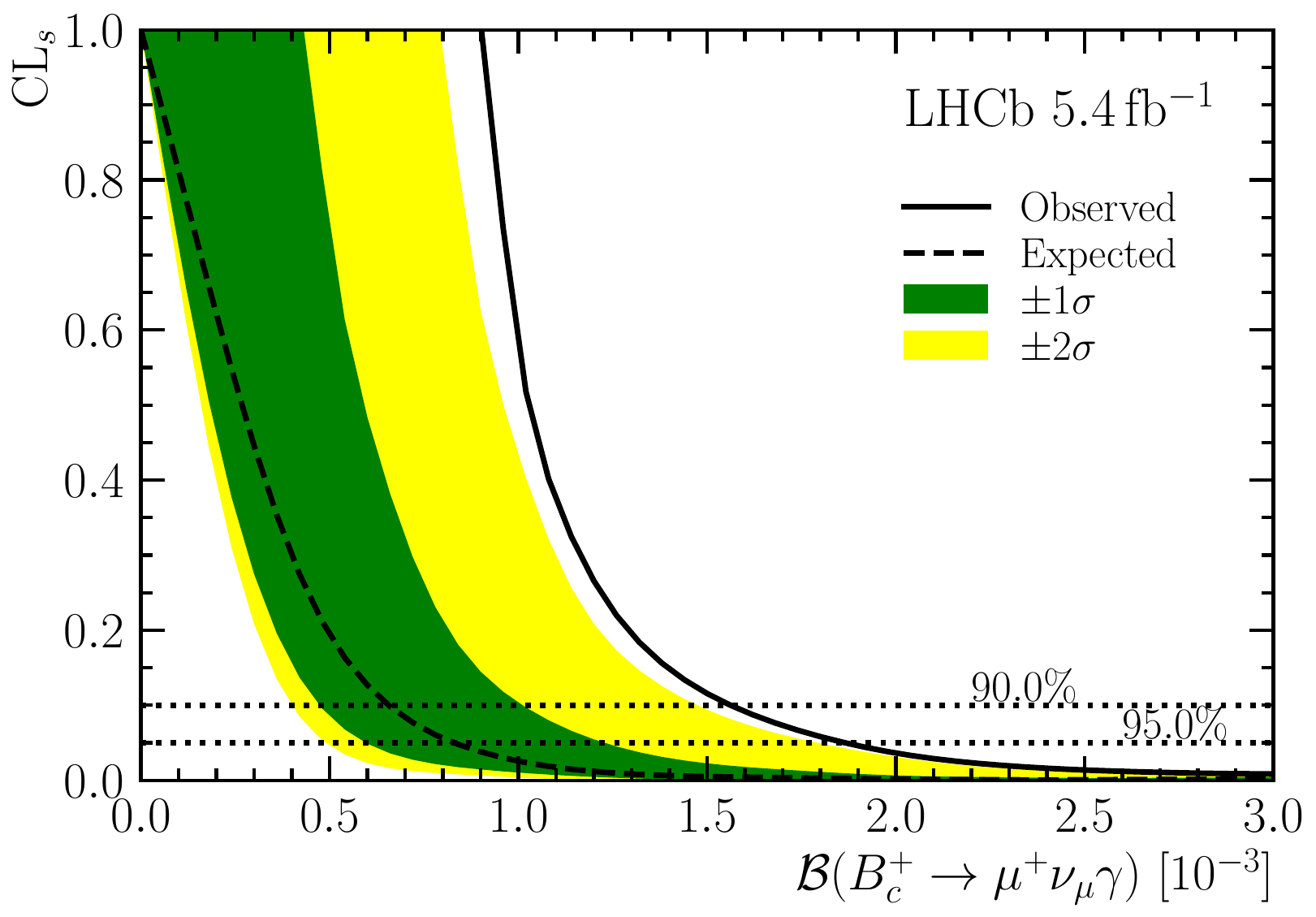}
    \caption{Results from the CL$_s$ scan used to obtain the limit on (left) $B^+\to\mu^+\nu_\mu\gamma$ and (right) $B_c^+\to\mu^+\nu_\mu\gamma$ branching fractions. The dashed line shows the expectation in the absence of signal and the $1\sigma$ and $2\sigma$ confidence intervals are shown as the green and yellow bands, respectively. The observation is shown as the solid black line. The two dotted horizontal lines intersecting with the observation indicate the limits at 90\% and 95\% CL, respectively.}
    \label{fig:limits}
\end{figure}

The profiles of the negative log-likelihood are shown in Fig.~\ref{fig:nll}.
The CL$_s$ curves from which the upper limits on the signal branching fractions are determined are shown in Fig.~\ref{fig:limits}.

\subsection*{Efficiency maps for reinterpretation of the results} 
\begin{figure}[h]
    \centering
    \includegraphics[width=0.45\linewidth]{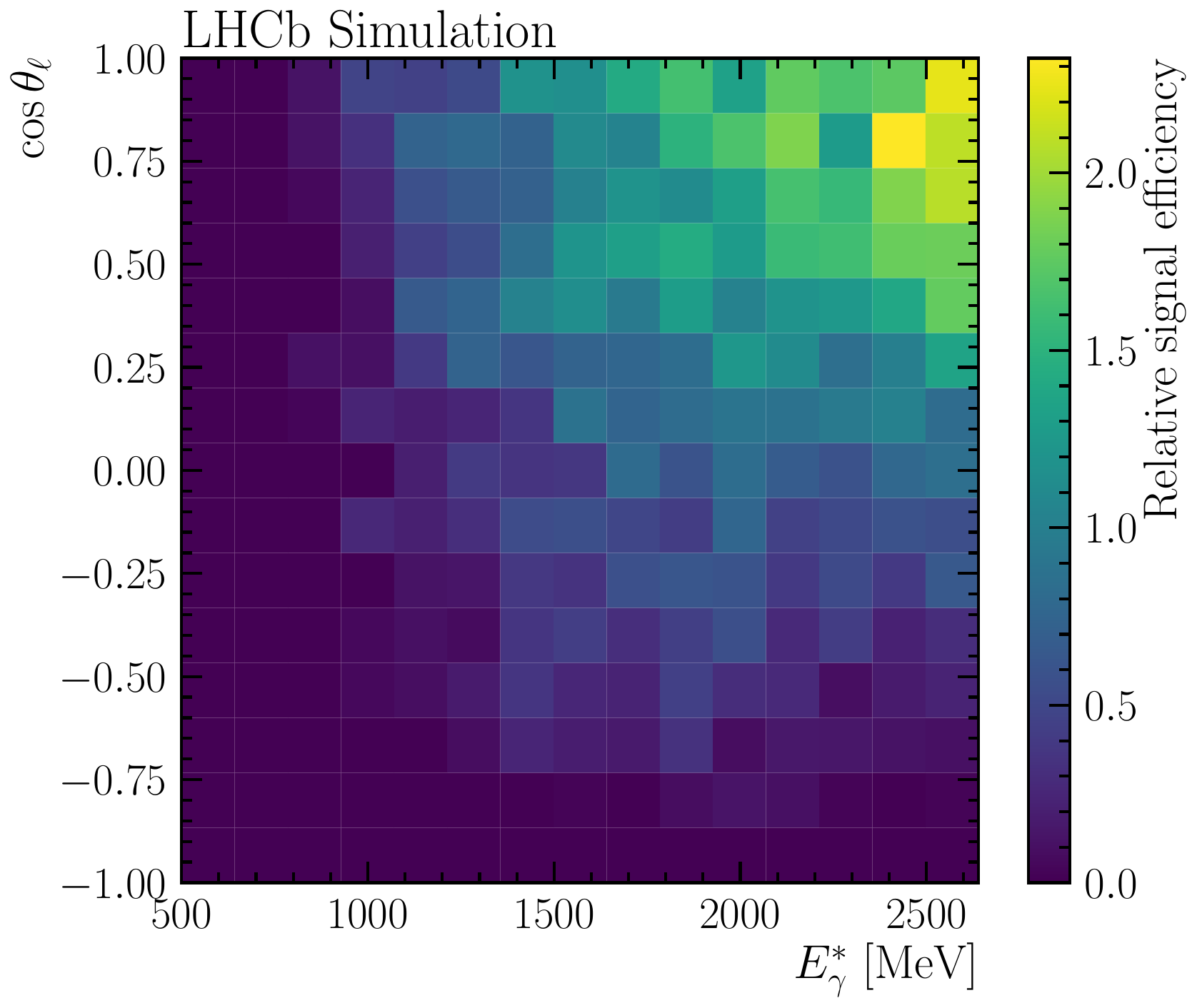}
    \includegraphics[width=0.45\linewidth]{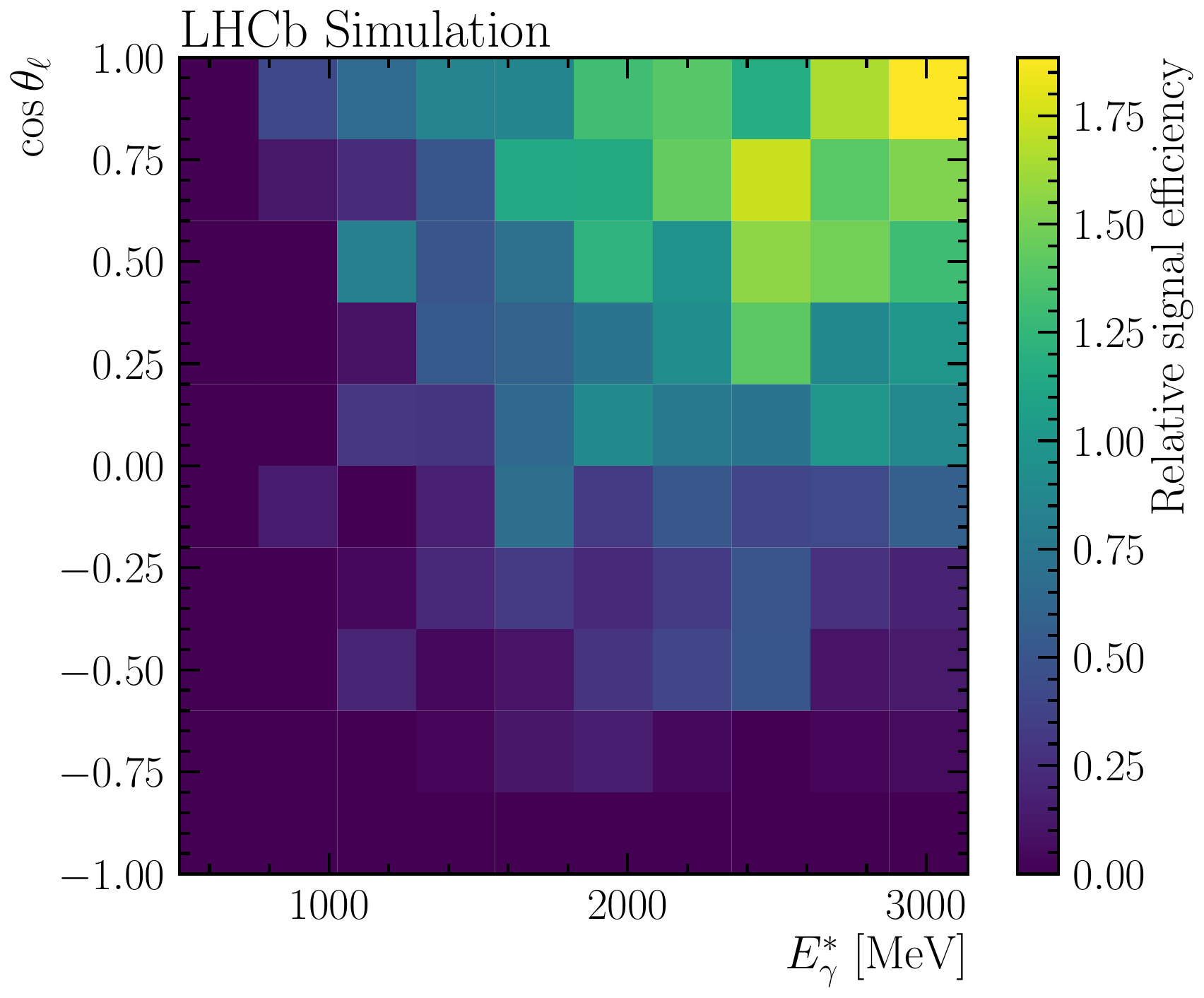}
    \caption{Two-dimensional map of the relative reconstruction and selection efficiency for (left) $B^+\to\mu^+\nu_\mu\gamma$ decays and (right) $B_c^+\to\mu^+\nu_\mu\gamma$ decays as a function of $E_\gamma^\ast$ and $\cos \theta_\ell$. The efficiencies are obtained from $B_{(c)}^+\to\mu^+\nu_\mu\gamma$ simulation samples generated with a phase-space model and are normalized to the integrated efficiency of the \texttt{LNuGamma} model.}
    \label{fig:model_eff2d}
\end{figure}

The \Bucmunugamma decay can be fully parametrized by the photon energy in the \B rest frame, \Egamma, and the helicity angle, $\theta_\ell$, defined in the $\mup\nu_\mu$ rest frame as the angle between the $\mup$ momentum direction and the opposite direction of the \B momentum, using the approximation from Ref.~\cite{LHCb-PAPER-2015-025}. 
The upper limits presented here are obtained using the model based on the leading-order calculation from Ref.~\cite{Korchemsky:1999qb}. The signal efficiency depends on the model due to the nonuniform selection efficiencies as a function of \Egamma and $\cos\theta_\ell$ that are shown in Fig.~\ref{fig:model_eff2d} for the \Bumunugamma and \Bcmunugamma decay modes. The digitized versions of these efficiency maps are provided in the HEPdata record~\cite{hepdata.180801} and allow a reinterpretation of the result within different models.  

\newpage
\addcontentsline{toc}{section}{References}
\bibliographystyle{LHCb/LHCb}
\bibliography{main,LHCb/standard,LHCb/LHCb-PAPER,LHCb/LHCb-CONF,LHCb/LHCb-DP,LHCb/LHCb-TDR}

\newpage
\centerline
{\large\bf LHCb collaboration}
\begin
{flushleft}
\small
R.~Aaij$^{39}$\lhcborcid{0000-0003-0533-1952},
M.~Abdelfatah$^{71}$,
A.S.W.~Abdelmotteleb$^{59}$\lhcborcid{0000-0001-7905-0542},
C.~Abellan~Beteta$^{53}$\lhcborcid{0009-0009-0869-6798},
F.~Abudin\'en$^{61}$\lhcborcid{0000-0002-6737-3528},
T.~Ackernley$^{63}$\lhcborcid{0000-0002-5951-3498},
A.A.~Adefisoye$^{71}$\lhcborcid{0000-0003-2448-1550},
B.~Adeva$^{49}$\lhcborcid{0000-0001-9756-3712},
M.~Adinolfi$^{57}$\lhcborcid{0000-0002-1326-1264},
P.~Adlarson$^{87,44}$\lhcborcid{0000-0001-6280-3851}
% ...

{\footnotesize \it

% ...
$^{39}$Nikhef National Institute for Subatomic Physics, Amsterdam, Netherlands\\
% ...
$^{44}$National Center for Nuclear Research (NCBJ), Warsaw, Poland\\
% ...
$^{49}$Instituto Galego de F{\'\i}sica de Altas Enerx{\'\i}as (IGFAE), Universidade de Santiago de Compostela, Santiago de Compostela, Spain\\
% ...
$^{53}$Physik-Institut, Universit{\"a}t Z{\"u}rich, Z{\"u}rich, Switzerland\\
% ...
$^{57}$H.H. Wills Physics Laboratory, University of Bristol, Bristol, United Kingdom\\
% ...
$^{59}$Department of Physics, University of Warwick, Coventry, United Kingdom\\
% ...
$^{61}$School of Physics and Astronomy, University of Edinburgh, Edinburgh, United Kingdom\\
$^{62}$School of Physics and Astronomy, University of Glasgow, Glasgow, United Kingdom\\
$^{63}$Oliver Lodge Laboratory, University of Liverpool, Liverpool, United Kingdom\\
% ...
$^{71}$Syracuse University, Syracuse, NY, United States\\
% ...
$^{87}$Department of Physics and Astronomy, Uppsala University, Uppsala, Sweden, associated to $^{62}$\\
% ...
}
\end{flushleft}

\end{document}